\documentclass{./elsart4-1}
\usepackage{amssymb,amsmath}
\usepackage{graphicx}
\usepackage{float}
\usepackage{bm}
\usepackage{multirow}
\usepackage{dcolumn}
\usepackage{tabularx}
\usepackage{tabulary}
\usepackage{longtable}
\usepackage[dvipsnames,usenames]{color}
\usepackage[english,francais]{babel}
\usepackage[T1]{fontenc}
\usepackage[latin1]{inputenc}

\newtheorem{e-proposition}[theorem]{Proposition}

\newtheorem{e-definition}[theorem]{Definition\rm}

\renewcommand{\theequation}{\arabic{equation}}
\begin{document}
\begin{frontmatter}

  \title{Novel magneto-electric multiferroics from first-principles}

\selectlanguage{english}
\author[JV]{Julien Varignon}
\ead{julien.varignon@ulg.ac.be}
\author[JV]{Nicholas C. Bristowe}
\ead{n.bristowe@ulg.ac.be}
\author[JV]{Eric Bousquet}
\ead{eric.bousquet@ulg.ac.be}
\author[JV]{Philippe Ghosez}
\ead{philippe.ghosez@ulg.ac.be}

\address[JV]{Physique Th\'eorique des Mat\'eriaux, Universit\'e de Li\`ege  (B5), B-4000 Li\`ege, Belgique}

\date{\today}

\medskip
\begin{center}
{\small Received *****; accepted after revision +++++}
\end{center}

\begin{abstract}
  Interest in first-principles calculations within the multiferroic community
  has been rapidly on the rise over the last decade. Initially considered as a
  powerful support to explain experimentally observed behaviours, the trend
  has evolved and, nowadays, density functional theory calculations has become
  also an essential predicting tool for identifying original rules to achieve
  multiferroism and design new magneto-electric compounds. This chapter aims
  to highlight the key advances in the field of multiferroics for which
  first-principles methods have contributed significantly. The essential
  theoretical developments that made this search possible are also briefly
  presented.

{\it To cite this article:
    J. Varignon, N.~C. Bristowe, E. Bousquet and Ph. Ghosez, C. R.  Physique
    \# (2014).}

\vskip 0.5\baselineskip

\selectlanguage{francais}
\noindent{\bf R\'esum\'e}
\vskip 0.5\baselineskip
\noindent {\bf Nouveaux mat\'eriaux multiferro\"{i}ques \`a partir de calculs de
  premiers principes. }

L'int\'er\^et pour les calculs {\em ab initio} dans la communaut\'e des
multiferro\"iques n'a cess\'e de cro\^itre au cours de la derni\`ere
d\'ecennie. Initialement consid\'er\'es comme un support efficace pour
expliquer les comportements observ\'es exp\'erimentalement, la tendance a
\'evolu\'e et, actuellement, les calculs r\'ealis\'es dans le cadre de la
th\'eorie de la fonctionnelle de la densit\'e apparaissent aussi comme un
outil pr\'edictif incontournable permettant d'identifier de nouvelles voies
pour parvenir au multiferro\"isme et cr\'eer de nouveaux mat\'eriaux
magn\'eto-\'electriques. Ce chapitre vise \`a pr\'esenter quelques avanc\'ees
clefs dans le domaine des multiferro\"iques, auxquelles les m\'ethodes {\it ab
  initio} ont conbribu\'e de mani\`ere significative.  Les d\'eveloppements
th\'eoriques essentiels ayant permis ces avanc\'ees sont aussi bri\`evement
discut\'es.

 {\it Pour citer cet article~: J. Varignon, N.~C. Bristowe, E. Bousquet and
Ph. Ghosez, C. R.  Physique \# (2014).}

 \vskip 0.5\baselineskip
\noindent{\small{\it Mots-cl\'es~:} DFT; multiferroique; magn\'etisme}}
\end{abstract}
\end{frontmatter}

\selectlanguage{english}

\section{Introduction}
\label{intro}

Discovered at the end of the 19$^{th}$ century, the magneto-electric effect in
which the magnetization can be tuned by an electric field and the polarization
by a magnetic field has seen significant developments during the '60-'70s but
remained at that time essentially an academic curiosity. A significant renewal
of interest for magneto-electrics has only appeared recently, during the
early~'00s~\cite{fiebig2005revival,eerenstein2006multiferroic}, boosted by
their potential for various technological applications
\cite{ME-memory,Kimura_Nature,Hur_Nature429,Goto_PRL92}. The field of
magneto-electrics is also intimately linked to that of multiferroics, although
not limited to them. It is indeed expected that the amplitude of the linear
magneto-electric (ME) effect, $\alpha$, is bound by the electric ($\chi^e$)
and magnetic ($\chi^m$) susceptibilities through the expression $\alpha^2 <
\chi^e \cdot \chi^m$~\cite{fiebig2005revival}. According to this, the
magneto-electric effect can {\em a priori} be very large in ferroelectrics
and/or ferromagnetics and therefore so-called multiferroic compounds combining
these properties have received focussed attention. It is not guaranteed
however that these will have the largest magneto-electric coupling.

A quite similar renewal of interest had appeared for ferroelectric compounds a
decade before, in the early '90s. At that time, prototypical ferroelectrics
such as BaTiO$_3$ and PbTiO$_3$ had become accessible to DFT
calculations~\cite{cohen1992origin}. Initially restricted to explain old
observations, the microscopic understanding acquired from first-principles
calculations rapidly enabled practical guidance for
experimentalists. Theoretical studies played a key role in clarifying the
microscopic origin of ferroelectric and piezoelectric
properties~\cite{cohen1992origin,vanderbilt1997first} and later ferroelectric
finite size
effects~\cite{junquera2003critical,dawber2005physics,Strain-review}. In the
00's, the interest for ferroelectrics then naturally extended to
magneto-electric multiferroics.

Over the last decade, first-principles calculations were particularly helpful
in the field of multiferroics. On the one hand, they provided microscopic
understanding of several experimental observations. They contributed
significantly to understand prototypical systems such as BiFeO$_3$
\cite{pairelibre}, YMnO$_3$ \cite{YMO-improper,YMO-VanAken} or TbMnO$_3$
\cite{TbMnO3-vanderbilt}. On the other hand, first-principles calculations
also appear to be a powerful design tool for making theoretical predictions,
eventually confirmed experimentally. It is this second aspect that is the
focus of the present Chapter.

Although not limited to them, many concrete advances to date in the field of
multiferroics naturally involved ABO$_3$ perovskites and related
compounds. Thanks to the wide variety of properties they can exhibit within
the same simple structure and the possibility to combine them in various
nanostructures, these compounds are providing a fantastic playground for both
theorists and experimentalists~\cite{zubko2011interface}. It was nevertheless
initially thought that ferroelectricity and magnetism are mutually exclusive
in this class of compounds : the apparent scarcity of ABO$_3$ multiferroics
was explained by the fact that their ferroelectric property is related to
O-$2p$-B-$3d$ hybridization and typically requires $d^0$ occupancy while
magnetism requires partial $d$-state filling \cite{d0rule}. The popular
room-temperature multiferroic BiFeO$_3$ (for a more complete discussion see
Chapter 3) circumvents this contradictory B-cation $3d$-filling
requirements~\cite{d0rule} with ferroelectricity and magnetism originating
from different A and B cations respectively. This has motivated the search for
related compounds such as Bi$_2$CrFeO$_6$ wich was predicted to be
multiferroic with a large polarization from first
principles~\cite{BFCO-Spaldin}, and later demonstrated experimentally on thin
films~\cite{BFCO-MF-exp1,BFCO-MF-exp2}. We will see here that $d^0$-ness is in
fact not always mandatory and that various strategies can be developed to
allow ferroelectricity and magnetism to coexist in ABO$_3$ compounds.

In this Chapter, we will first briefly describe in Section 2 the essential
theoretical advances that were required to make density functional theory
calculations predictive and have fuelled theoretical discoveries in the field
of multiferroics. Without being exhaustive, we will then present in the next
Sections selected strategies that have been proposed to achieve multiferroism
and in which theory has played a central role. Although at first glance quite
distinct, we will see that many of these approaches finally rely on a common
concept: how to make a paraelectric magnet ferroelectric.  This can be
achieved either by strain engineering (Section 3), lattice mode engineering
(Section 4) or electronic spin, orbital or charge engineering (Section 5).  In
all cases a more or less direct magneto-electric coupling is realized.  We
will briefly address the case of magnetic/ferroelectric interfaces in Section
6, before concluding in Section 7.

We note that the prototypical BiFeO$_3$ system has been extensively studied in
the multiferroic community. Due to the volume of work on this material, we
will only present a selection of key results coming from first principles and
we refer the reader to the dedicated chapter within this issue for a more
complete discussion on this compound.

\section{First-principles density functional theory methods}
\label{methode}

Density functional theory (DFT) was proposed during the mid-sixties. Groun\-ded
in the Hohenberg and Kohn theorem \cite{HK-theorems} and the Kohn and Sham
ansatz \cite{KohnSham}, it was initially a purely theoretical concept that
remained dormant for almost 20 years, until the advent of efficient computers
enabled the transformation into a very powerful computational method. Since
the eighties, DFT has seen an explosive growth, driven both by the ongoing
increase of computer power and various concomitant theoretical and algorithm
developments (see for instance the textbook of R. M. Martin for a
comprehensive description of DFT ~\cite{martin2004electronic}). Although {\it a
  priori} an exact theory, the practical implementation of DFT relies on
approximations giving rise to well-known limitations (see Section
\ref{AM}). Nevertheless, the method has proven to be an excellent compromise
between accuracy and efficiency.  Nowadays, it has become an essential
approach in materials research. Aside from the Nobel prize in Chemistry
attributed to W. Kohn in 1998 for this specific contribution, it is worth
noticing that amongst the 10 most-cited papers of Physical Review journals, 6
are directly related to density functional theory.

First-principles DFT calculations have certainly contributed to the revival of
interest for ferroelectrics in the early nineties. The theoretical study of
ferroelectrics took advantage of density functional perturbation theory
\cite{Baroni-87,Gonze-92} to access systematically by linear response various
dynamical and piezoelectric properties \cite{Gonze-97,RMP-Baroni} and,
reciprocally, contributed to further develop it \cite{Wu-05}.  It also boosted
the discovery of the ``modern theory of polarization'' by King-Smith and
Vanderbilt~\cite{KSV,VKS} and Resta \cite{Resta-1,Resta-2}. This fundamental
breakthrough allows to clarify the fundamental role of polarization in
periodic DFT~\cite{gonze1995density} and further gave rise to numerous
advances, such as the development of finite electric and displacement field
techniques~\cite{sai2002theory,souza2002first,stengel2009electric}.

Addressing the physics of magneto-electric multiferroics required to face
additional new challenges.  On the one hand the modelling of magnetic systems
is intrinsically much more demanding computationally and methods had also to
be developed to access the magneto-electric coefficients. On the other hand,
magneto-electric multiferroics are typically strongly-correlated systems for
which the usual local density approximation (LDA) or generalized gradient
approximation (GGA) to DFT are not appropriate, so that alternative more
advanced methods had to be envisaged.

\subsection{Magnetic systems}
\label{MS}

In magnetic compounds, both the spin and the orbital motion of the electrons
contribute to the total magnetization. While the spin contribution to the
magnetization of periodic solid has been accessible from DFT methods for many
years, the way to compute the orbital one has only been formulated recently.

Kohn-Sham DFT can be at first trivially extended to spin-polarized systems by
simply treating separately the density of up and down
spins~\cite{martin2004electronic}. This constitutes a {\it collinear-spin}
level of approximation in which the magnetic moment appears as a scalar
quantity. Although widely used, this is not however the most general
formulation since the spin axis can vary in space.  Extension of DFT to the
{\it non-collinear spin} level was first formulated by von Barth and
Hedin~\cite{Barth-Hedin}.  Here the density is no longer a scalar but a $2
\times 2$ matrix $\overline{\overline{n}}(\vec r)$ depending on the scalar
density $\rho(\vec r)$ and the magnetic density $\vec m (\vec r$) :
\begin{equation}
\label{eq:density-noncol2}
\overline{\overline{n}}(\vec r) = \frac{1}{2}\left( \rho(\vec r)\cdot
  \overline{\overline{\rm I}} +
  \sum_{i=x,y,z} m_i(\vec r)\cdot\overline{\overline{\sigma}}_i  \right)
\end{equation}
\begin{equation}
\label{eq:density-noncol}
\overline{\overline{n}}(\vec r) \,\, \Rightarrow \,\, \frac{1}{2}
\left(
\begin{array}{cc}
  \rho(\vec r) +m_z(\vec r) & m_x(\vec r) -i m_y(\vec r) \\
  m_x(\vec r) +i m_y(\vec r) & \rho(\vec r) -m_z(\vec r)
\end{array}
\right)
\end{equation}
where $\overline{\overline{\sigma}}$ are the Pauli matrices. The spin-density
matrix $\overline{\overline{n}}(\vec r)$ now allows the magnetization to relax
in direction and magnitude, providing access to non-collinear magnetic
structures. At this level the coupling between the spins and the lattice has
to be explicitly included through the spin-orbit interaction.

Most investigations to date make use of the collinear spin approximation.
Although the formalism is well known, non-collinear spin calculations on
concrete systems of interest remain very challenging since the energy scale
involved is typically extremely small.  Calculations require a high degree of
convergence and the search for the ground-state spin configuration is
complicated by the existence of numerous local minima. Moreover, as further
discussed later (see Section \ref{AM}), the final result is often sensitive to
the chosen approximations and hence must always be considered with care.

As for the electric polarization, computation of the orbital magnetization in
periodic systems remained elusive for many years. The problem has only been
solved recently~\cite{MTM,BerryPhase-magnetization}, providing an expression
that can be easily implemented in DFT codes~\cite{Mauri-10} . This can be seen
as a Berry phase analogue of the theory of polarization. We refer the reader
to Ref.~\cite{Resta-JPCM,malashevich2010theory} for a more complete discussion
of the so-called modern theory of magnetization.

\subsection{Computing the magnetoelectric coefficients}
\label{ME}

The magneto-electric tensor $\alpha$ is a mixed second derivative of the
energy ($\mathcal{F}$) that describes the change of magnetization ($M$)
produced, at linear order, by an electric field ($E$) or, equivalently, of
polarization ($P$) produced by a magnetic field ($H$):
\begin{equation}
\label{alpha_ij}
\alpha_{ij} = \frac{-1}{\Omega_0} \frac{\partial^2 \mathcal{F}}{\partial E_i \partial H_j} = \frac{\partial P_i}{\partial H_j} =  \frac{\partial M_j}{\partial E_i} 
\end{equation}
where $\Omega_0$ is the unit cell volume. Considering electronic, ionic and
strain degrees of freedom as independent parameters (Born-Oppenheimer
approximation), $\alpha$ can be conveniently decomposed into 3 terms :
\begin{equation}
\label{alpha}
\alpha = \alpha^{el} + \alpha^{ion} + \alpha^{strain}
\end{equation}
where $\alpha^{el}$ is the purely electronic response (at fixed geometry),
$\alpha^{ion}$ the additional contribution coming from the ionic relaxation
and $\alpha^{strain}$ the additional contribution coming from the strain
relaxation. Keeping in mind that the magnetization can have a spin ($S$) and
orbital ($O$) origin, $\alpha$ can be viewed as consisting of six individual
contributions (see table~\ref{tab:MEdecompo}). Although a key quantity in the
study of magneto-electrics, it is worth noticing that the methods providing
access to these different terms have only been made accessible recently.
\begin{table}[h!]
\centering
\begin{tabular}{c|c|c|c}
  \hline
  \hline
  & Electronic & Ionic   & Strain \\
  \hline
  Spin     &  $\alpha^{el}_S$ (2011~\cite{Eric}) & $\alpha^{ion}_S$ (2008~\cite{MethodeJorge}) & $\alpha^{strain}_S$ (2009~\cite{ME-Wojdel1})       \\
  Orbital  &  $\alpha^{el}_O$  (2012~\cite{PhysRevB.86.094430})   &
  $\alpha^{ion}_O$ (2012~\cite{PhysRevB.86.094430,Eric-orbLiFePO4}) & $\alpha^{strain}_O$ (--) \\
  \hline
  \hline
\end{tabular}
\caption{Individual contribution to the linear magnetoelectric coupling tensor
  $\alpha$. The year it was first computed and the relevant reference are
  provided in brackets.}
\label{tab:MEdecompo}
\end{table}

Pioneering computations of the linear magneto-electric coefficients have been
performed by I\~niguez {\em et al}. Assuming a dominant $\alpha^{ion}_S$
contribution, I\~niguez \cite{MethodeJorge} proposed a scheme to access in a
linear response framework the change of spin magnetization resulting from the
ionic relaxation produced by an electric field, in a similar spirit to what is
usually done to access the ionic contribution to the dielectric
constant~\cite{Gonze-97}. The method was then naturally extended to the strain
contribution by Wojdel and I\~niguez~\cite{ME-Wojdel1}. In their derivations,
the last two terms of Eq.~\ref{alpha} take the form :
\begin{equation}
\label{Eq-Iniguez}
\alpha_{S,ij} = \alpha_{S,ij}^{el} +  \frac{1}{\Omega_0} \sum_{n=1}^{N_{IR}}
\frac{p_{ni}^{d}\,p_{nj}^{m}}{K_n} +  \sum_{m,n=1}^{6} e_{im} (C^{-1})_{mn} h_{jn} 
\end{equation}
 The ionic contribution (second term) involves a sum over the IR active
modes; it is directly proportional to the mode dielectric polarity
($p_{ni}^{d} = \sum_{at} Z^{*}_{at,i} u_{ni}$, where $Z^{*}_{at}$ is the Born
effective charge tensor and $u_{n}$ the phonon eigenvector) and the magnetic
equivalent ($p_{ni}^{m} = \sum_{at} Z^{m}_{at,i} u_{ni}$, where $Z^{m}_{at}$
is the magnetic effective charge tensor~\cite{PhysRevB.89.064301}) and is
inversely proportional to the force constant eigenvalues ($K_n$). The strain
contribution (third term) involves the piezoelectric ($e_{im}$) and
piezomagnetic ($h_{jn}$) constants and the inverse of the elastic
  constants matrix ($(C^{-1})_{mn}$).  Except for $p_{nj}^m$ and $h_{jn}$,
most of the quantities appearing in Eq. \ref{Eq-Iniguez} were already
routinely accessible by linear response or finite difference
techniques~\cite{RMP-Baroni}. In their work, $p_{nj}^m$ and $h_{jn}$ were
determined from finite differences. Eq. \ref{Eq-Iniguez} provides insight on
the microscopic origin of $\alpha^{ion}_S$ and $\alpha^{strain}_S$, suggesting
a route to design a large contribution: as proposed in Ref. ~\cite{ME-Wojdel2}
engineering structural ``softness'' (vanishingly small eigenvalues of {\bf$C$}
or {\bf$K$}) will produce a diverging behavior.

As an alternative to the previous linear-response approach, Bousquet {\em et
  al}~\cite{Eric} proposed to access the magneto-electric coefficients from
calculations of the change of macroscopic polarization in a finite magnetic
field.  In their work, the authors proposed to include the effect of the
magnetic field through adding a Zeeman term $\overline{\overline{\Delta V}}_{Zeeman}$
(applied on spins only) to the external potential $\hat{V}_{ext}$ with the
following expression in the $2 \times 2$ representation for non-collinear
magnetism:
\begin{equation}
\label{eq:Zeeman}
\overline{\overline{\Delta V}}_{Zeeman} = 
\frac{-g}{2}\mu_B\mu_0\left(
\begin{array}{cc}
H_z & \,\,\,\,\,\,H_x +iH_y \\
H_x -iH_y &\,\,\,\,\,\, -H_z
\end{array}
\right)
\end{equation}
where $\vec H$ is the applied magnetic field. The magneto-electric
coefficients are then deduced from calculations at different amplitudes of the
field by finite difference : $\alpha_{S,ij} = \Delta P_i / \Delta H_j$. On the
one hand, calculations in finite $H$ field at fixed ionic positions and
strains have given access for the first time to $\alpha_S^{el}$. On the other
hand, calculations including structural relaxation provide alternative access
to $\alpha^{ion}_S$ and $\alpha^{strain}_S$. Although in multiferroics
$\alpha^{ion}_S$ is expected to dominate especially around the ferroelectric
phase transition, Bousquet \textit{et al.} have shown that in the
magneto-electric Cr$_2$O$_3$, $\alpha_S^{el}$ is comparable in magnitude to
$\alpha_S^{ion}$ and therefore by no means negligible. It is worth noticing
also that such a finite field approach is not restricted to the determination
of $\alpha$ but also the higher-order responses.

The calculation of the orbital magnetic response came slightly later, with the
emergence of the modern theory of magnetization ~\cite{OrbME1,OrbME2,OrbME3}.
Using this technique, Malashevich \textit{et al.}~\cite{PhysRevB.86.094430}
computed $\alpha_O^{el}$ and $\alpha_O^{ion}$ for Cr$_2$O$_3$ from the change
of $M$ in a $E$ field and they indeed confirmed that the orbital contribution
is much smaller than the spin one. At the same time, Scaramucci \textit{et
  al.}~\cite{Eric-orbLiFePO4} computed $\alpha_O^{ion}$ in LiFePO$_4$. They
used the approximation of integrating the orbital moment with spheres centred
on each atom instead of the exact modern theory treatment and studied
$\alpha_O^{ion}$ using a method similar to that of I\~{n}iguez
\cite{MethodeJorge}. Interestingly, these results show that $\alpha_O^{ion}$
in LiFePO$_4$ is as large as $\alpha_S^{ion}$ and is even as large as the full
ME response of Cr$_2$O$_3$.

Alternative methods have been designed to overcome the fact that only the spin
contribution at 0~K is taken into account in the calculation of the ME
response within DFT.  As the temperature increases, spin fluctuations arise
and can induce an additional contribution to $\alpha^{spin}$. This is the so
called exchange-striction mechanism.  Mostovoy {\em et al} developed a method
to take into account this temperature effect by combining Monte-Carlo
simulations on a Heisenberg-type Hamiltonian in which the exchange parameters
were calculated from DFT calculations~\cite{Mostovoy-MET}. They applied their
method on Cr$_2$O$_3$ and showed that the exchange-striction mechanism can
induce a non-zero and large ME response along a direction that would be zero
otherwise. The spin fluctuations break the inversion center and induce a
polarization in the crystal. They also showed that the spin-orbit origin of
the ME response is one order of magnitude smaller than the exchange-striction
contribution when it reaches its maximum at a given temperature.  

The linear and non linear magneto-electric coefficient at finite temperature,
and the origin of the spin spiral of BiFeO$_3$ have also been calculated in
the framework of an effective
hamiltonian~\cite{PhysRevB.83.020102,Bellaiche-cycloidBFO}

\subsection{Beyond LDA and GGA}
\label{AM}

By default, DFT calculations are performed within the so-called Local Density
Approximation (LDA) or Generalized Gradient Approximations (GGA).  Although
these have proven to be highly predictive for many classes of compounds such
as band insulators and simple metals, they fail to describe systems with
strong electronic correlations (see Ref.~\cite{martin2004electronic}).
 
Since multiferroics involve correlated systems, more advanced functionals are
required to capture the basic physics. The most simple and popular approach is
the LDA+U method which involves two empirical parameters U and J, accounting
for the on-site Coulomb interactions and the intra-site spin exchange
respectively~\cite{LDAU}.  Both are captured in an effective way through a
Hubbard-like model. Two different implementations of LDA+U are commonly used:
one adopting two independent parameters~\cite{LDAU-Lich} and the other using
only one effective parameter $\Delta U_{eff}=$U-J~\cite{LDAU-Duda}. In either
case, U and J (or $\Delta U_{eff}$) are adjustable parameters and one must fit
their value in order to reproduce experimental trends. Alternatively, a self
consistent method to calculate the parameters exists~\cite{SelfC-U}, however
it does not always appear to be fully predictive and the parameters sometimes
have to be rescaled~\cite{BFO-Bellaiche-FiniteT}. In practice, the basic
properties of the material are extremely sensitive to the values of U and J.
While at the spin collinear level, J is commonly neglected, this parameter
becomes important and meaningful when going to non collinear spin calculations
as it acts on on the off diagonal terms of equation~\ref{eq:density-noncol},
in other words on the spin canting of the system~\cite{Eric-J}.

In order to overcome the adjustable parameters of LDA+U, hybrid functionals
are a valuable alternative method which have become widely used
nowadays~\cite{BFO-Marco,Hybrid-Picozzi,BMO-improper,Alina-YMO,Hybrid-Picozzi-Vanderbilt}. These
functionals take their name from consisting of a combination of LDA and GGA
functionals plus a part of exact exchange to reproduce exchange and
correlation effects more accurately. The most famous hybrid functionals are
the B3LYP~\cite{B3LYP} and HSE~\cite{HSE}. A B1WC functional has also been
optimized for ABO$_3$ ferroelectrics~\cite{B1WC} and revealed powerful for
multiferroics as well~\cite{BFO-Marco,BMO-improper,Alina-YMO}. Unfortunately,
these hybrid functionals are not widely implemented in DFT codes, and
additionally are more computationally expensive. Consequently, some groups are
then using them as a benchmark to extract the adjustable parameters for a more
computationally tractable LDA+U calculation~\cite{Hybrid-Picozzi-Vanderbilt}.

While LDA+U and hybrid functionals are the most commonly used approach within
the field of multiferroics, several alternative methods exist. In order to
remedy failures in DFT arising from the spurious self interaction term
(interaction of an electron with the potential generated by itself),
Filippetti and Spaldin proposed a method to better approximate the correction
to this term within pure LDA calculations, involving minimal computational
costs (pseudo-Self Interaction Correction method)~\cite{PseudoSIC}.  Dynamical
Mean Field Theory (DMFT) is an alternative method to describe correlation
effects by going beyond the static mean field theory used in
DFT~\cite{DMFT1,DMFT2}. Alternatively, a quantum chemistry method has been
developed to accurately evaluate magnetic couplings in strongly correlated
systems~\cite{SAS} and has been used to study the evolution of the magnetic
exchange integrals with an external electric field~\cite{JVYMnO3}. The most
accurate but expensive parameter free theoretical method including many body
effects is the GW method, which has been used as a benchmark for DFT
calculations on BiFeO$_3$~\cite{Hybrid-Picozzi}.

We conclude the section by emphasizing that DFT calculations are restricted to
0~Kelvin in practice. In the field of ferroelectrics, this limitation has been
elegantly overcome through the developments of a so-called effective
hamiltonian method as pioneered by Zhong, Rabe and
Vanderbilt~\cite{EffectiveH-VDB-Rabe,EffectiveH-VDB-Rabe2}. This method has
been generalized by Bellaiche and co-workers for the case of
multiferroics~\cite{BFO-Bellaiche-FiniteT}. So far it has been applied to
BiFeO$_3$ yielding many key
advances~\cite{Bellaiche-EffectiveH-BFO,Bellaiche-cycloidBFO,PhysRevLett.112.147601,PhysRevLett.105.057601,PhysRevLett.103.047204}.


\section{Strain engineering}
\label{strain}

During the early 2000's, much effort was devoted to the understanding of the
role of electrical and mechanical boundary conditions on the ferroelectric
properties of ABO$_3$ perovskite thin films
\cite{dawber2005physics,Strain-review}. In 2004, it was shown, for instance,
that SrTiO$_3$, which is paraelectric at the bulk level, can be made
ferroelectric in thin film form and develop a spontaneous polarization at room
temperature under moderate epitaxial tensile strain~\cite{SrTiO3-FERT}.  The
idea naturally emerged to apply a similar strategy to turn paraelectric
magnets into ferroelectrics and make them {\it de facto} multiferroics.

\subsection{Inducing ferroelectricity by strain in magnetic systems}

\begin{figure}
\centering
\resizebox{12cm}{!}{\includegraphics{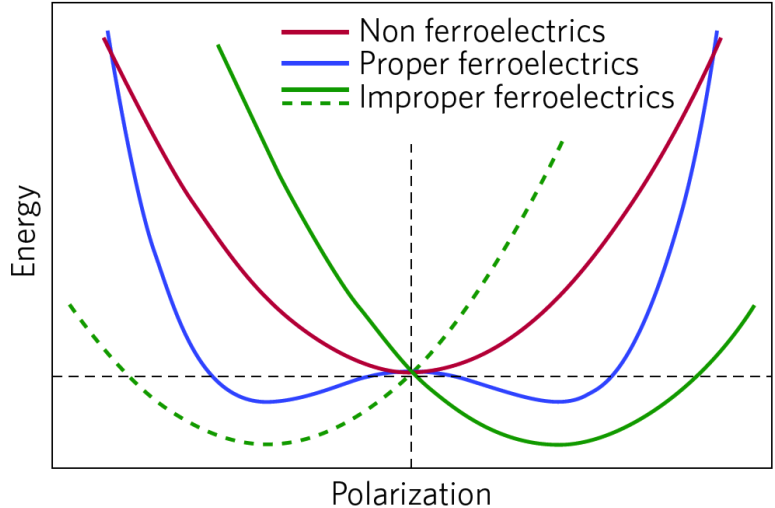}}
\caption{Energy potential as a function of the polarization. Figure taken from
  Ref.~\cite{nature-Ph}.}
\label{f:potential}
\end{figure}

Strain engineering of ferroelectricity is quite a universal approach, based on
polarization ($P$) -- strain ($\eta$) coupling. In simple cubic perovskites,
this $P$--$\eta$ coupling contributes to the Landau free energy ${\mathcal F}$
through a term of the form (at the lowest order):
\begin{equation}
\label{P-Strain}
 {\mathcal F} (P,\eta) \approx g  \,\, \eta P^2
\end{equation}
In non-ferroelectric compounds, the curvature of the energy respect to the
polarization is positive at the origin (red curve in
Fig.~\ref{f:potential}). From Eq. \ref{P-Strain}, it appears that one effect
of the strain is to renormalize this energy curvature. When producing a
sufficiently large negative contribution, the polarization-strain coupling can
hence destabilize the system and make it a proper ferroelectric (blue curve in
Fig.~\ref{f:potential}). We notice that turning the system into ferroelectric
is {\it a priori} possible, whatever the sign of the electro-strictive
coefficients $g$, through an appropriate choice of the strain (compressive or
tensile). In practice, the feasibility of the approach relies however on the
amplitude of the requested strain. So, starting from compounds on the verge of
ferroelectricity is clearly an asset.

As a concrete illustration of strain-induced ferroelectricity in magnetic
systems, let us consider CaMnO$_3$, a well-known G-type anti-ferromagnetic
(AFM-G) insulator. At the bulk level, CaMnO$_3$ exhibits a paraelectric $Pnma$
orthorhombic structural ground state, consisting of a slight distortion of the
ideal perovskite structure produced by antiferrodistortive (AFD) oxygen
motions.  Using first-principles calculations, Bhattacharjee {\it et al.}
~\cite{CaMnO3-Ph} have shown that, in its cubic phase, CaMnO$_3$ does in fact
combine weak ferroelectric (FE) and strong AFD instabilities.  Although the
former is suppressed by the appearance of the oxygen motions in the $Pnma$
phase, they predicted that orthorhombic CaMnO$_3$ can be made ferroelectric
under moderate epitaxial tensile strain, which was subsequently confirmed
experimentally~\cite{CaMnO3-Prellier}. Additionally, they also pointed out
that the FE distortion in CaMnO$_3$ is dominated by the Mn motion,
demonstrating that, in contrast to the previously discussed $d^0$
rule~\cite{d0rule}, the same cation can be responsible for the magnetic and
ferroelectric properties.  

Inducing ferroelectricity by strain in a magnetic compound such as CaMnO$_3$
makes it multiferroic but does not necessarily guarantee strong
magneto-electric coupling.  Bousquet and Spaldin~\cite{Eric-ME-CaMnO3}
nevertheless highlighted in 2011 that the appearance of a spontaneous
polarization in $Pnma$ perovskites can also give rise to a linear
magneto-electric effect. In the $Pnma$ phase, the symmetry allows for a small
canting of the otherwise anti-ferromagnetically ordered spins, yielding weak
ferromagnetism~\cite{bellaiche2012simple}.  A center of inversion is preserved
however in which case weak ferromagnetism is incompatible with a linear
magneto-electric effect \cite{FENNIE}. Inducing ferroelectricity by strain
breaks the inversion symmetry and additionally offers the possibility of
achieving a linear magneto-electric coupling. In the resulting ferroelectric
$Pmc2_1$ phase adopted by CaMnO$_3$ under moderate epitaxial tensile strain,
Bousquet and Spaldin predicted a linear magneto-electric coefficient much
larger than that of more conventional magneto-electrics like
Cr$_2$O$_3$. Their analysis, based on symmetry arguments, is totally
general. Since the $Pnma$ structure is the most common ground state in ABX$_3$
compounds, this finding generates a tremendous number of possibilities for
creating new magneto-electric materials under epitaxial strain.

Although not related to strain engineering, we notice here that the interplay
between ferroelectric distortion and weak ferromagnetism had also been
discussed independently by Fennie \cite{Fennie-08}.  The author proposed
design rules for identifying compounds in which ferroelectric distortions can
induce weak-ferromagnetism. In such cases, the weak magnetic moment is
directly proportional to P: switching one will automatically switch the other,
therefore opening the door to electric switching of the magnetization.  Using
first-principles calculations, he proposed $R3c$ FeTiO$_3$ and related
compounds as a promising realization of these ideas.  It was further confirmed
experimentally that the $R3c$ phase of FeTiO$_3$ is indeed ferroelectric and a
weak ferromagnet \cite{Varga}.

 Strain engineering can give rise to other unexpected phases in
  bulk perovskites such as the supertetragonal T phase (under
  compression)~\cite{Béa-SuperTphase,Jorge-TPhase} or the $Pmc2_1$ phase
  (under tension) of highly-strained
  BiFeO$_3$~\cite{PhysRevLett.109.057602,PhysRevLett.112.057202} and in
superlattices such as the ferromagnetic-ferroelectric $Pc$ phase of
SrTiO$_3$/SrCoO$_3$~\cite{song2014first}. Besides ABO$_3$ perovskites, simple
magnetic AO binary oxides surprisingly also appear as promising candidates to
multiferroism through epitaxial strain. Bousquet {\em et al.}~\cite{BaO-EuO}
proposed that EuO can be made ferroelectric under experimentally achievable
epitaxial strains.  Combined with the ferromagnetic character of EuO, this
would open interesting new perspectives if experimentally verified.

Moreover, strain engineering is not restricted to the possibility of inducing
ferroelectricity in paraelectric magnets. It can also be used to monitor the
properties of regular magnetoelectrics. From their calculations Wojdel and
I\~{n}iguez have shown the possibility to achieve a large enhancement of the
linear magnetoelectric coupling by inducing ``structural softness'' (see
section~\ref{ME}) through epitaxial strain in BiFeO$_3$
films~\cite{ME-Wojdel2}. In that study, they exploit the fact that, under
compressive epitaxial strain BiFeO$_3$ exhibits a structural phase transition
from a ferroelectric rhombohedral to a ferroelectric supertetragonal T
phase. Close to the critical strain, there is a region where the compound
becomes structurally soft and displays large responses.  Strain can
  further be used to tune the magnetic properties of the T phase of
  BiFeO$_3$~\cite{PhysRevLett.109.247202}.  Another example where strain is
used to tune the competition between alternative ordered phases is provided in
the next section.

\subsection{Exploiting large spin-lattice coupling}

In magnetic systems, the optical modes at the zone-center, consisting of
relative motions of distinct sublattices, are able to affect spin-spin
couplings.  The frequency $\omega$ of the zone-center optical modes are
therefore expected to be particularly dependent of the spin arrangement.  This
is the so-called ``spin-lattice'' coupling that can be approximated as
\cite{Baltensperger-70,Sabiryanov-99,Fennie-06} :
\begin{equation}
\label{sl}
 \omega \approx \omega_{PM} + \gamma <{\bold S}_i \cdot {\bold S}_j>
\end{equation}
where $\omega_{PM}$ is the frequency in the paramagnetic state , $\gamma$ is
the spin-lattice coupling constant and $<{\bold S}_i \cdot {\bold S}_j>$ the
nearest-neighbour spin-spin correlation function. This coupling in fact gives
rise to a non-linear magneto-electric effect, whatever the symmetry of the
magnetic system. In displacive ferroelectrics, the ferroelectric phase
transition is driven by the softening of a polar zone-center mode that
condenses at the phase transition producing a finite spontaneous polarization
\cite{Rabe-07}. When $\gamma$ is substantial, the spin-lattice coupling will
produce additional tuning of the soft-mode. In a simple Landau expansion of
the free energy, the essential physics of this tuning behavior can be included
through a bi-quadratic term~\cite{EutiO3-Exp} :
\begin{equation}
  \label{spin-lattice}
  {\mathcal F}(P,M) \approx - \gamma'   \,\, M^2 P^2 \,\,\,\,\,\,\,\,\,\, {\mathcal F}(L,M) \approx + \gamma'   \,\, L^2 P^2
\end{equation}
where $M$ and $L$ are the ferromagnetic and antiferromagnetic order
parameters. As for the strain in Eq. \ref{P-Strain}, we see that the magnetic
order is able to renormalize the curvature of the energy versus polarization
curve.  By tuning the system through an external parameter such as strain, the
critical value of the parameter that can make the system ferroelectric will
hence depend on the magnetic configuration.
   
\begin{figure}
\centering
\resizebox{12cm}{!}{\includegraphics{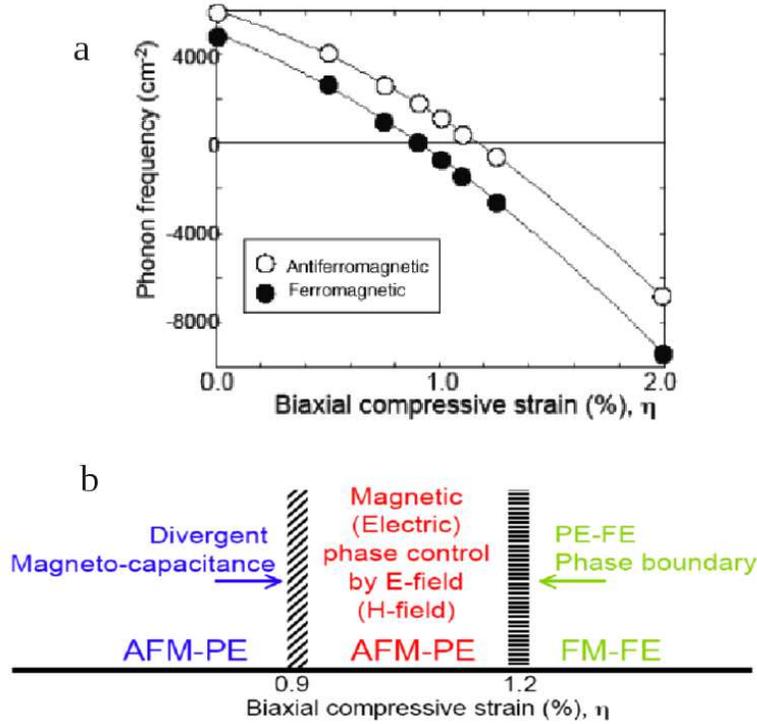}}
\caption{a) Evolution of the polar frequency $\omega²$ as a function of the
  compressive strain $\eta$ within a ferromagnetic (filled circles) and
  antiferromagnetic (unfilled circles) ordering. b) Phase diagram of EuTiO$_3$
  as a function of the compressive strain $\eta$. Figures reprinted with
  permissions from taken from C. J. Fennie, K. M. Rabe, Phys. Rev. Lett. 97
  (2006) 267602. (Ref.~\cite{Fennie-06}). Copyright (2014) by the American
  Physical Society.}
\label{f:EuTiO3}
\end{figure}

Using first-principles calculations, Fennie and Rabe \cite{Fennie-06} proposed
to exploit this effect in EuTiO$_3$, an antiferromagnetic insulating
perovskite that, in bulk, remains cubic and paraelectric at all
temperatures. This system exhibits a large and positive spin-lattice coupling
that further stabilizes the paraelectric state in the AFM configuration
(positive term in Eq. \ref{spin-lattice}) over a hypothetical FM state
(negative term in Eq. \ref{spin-lattice}). Exploiting strain to induce
ferroelectricity in EuTiO$_3$ (compressive or tensile strain can be used due
to the opposite $g$ coefficient associated to in-plane and out-of-plane $P$ in
Eq. \ref{P-Strain}) the amplitude of the critical strain needed to make the
system ferroelectric will be smaller for the FM state ($\eta_{c}^{FM}$) than
for the AFM one ($\eta_{c}^{AFM}$) (figure~\ref{f:EuTiO3}.a). They
demonstrated that in the intermediate strain region $ | \eta_{c}^{FM} | < |
\eta | < | \eta_{c}^{AFM} | $ the system offers easy magnetic control of the
polarization and {\it vice-versa}: in this region the system is still an AFM
paraelectric and aligning the spins in a magnetic field (resp. inducing a
polarization with an electric field) will induce a substantial polarization
(resp.  magnetization) by switching the system to the alternative FM
ferroelectric state (figure~\ref{f:EuTiO3}.b). Going further they highlighted
that under sufficiently large epitaxial strain the system will even adopt a FM
ferroelectric ground-state. This was further confirmed
experimentally~\cite{EutiO3-Exp} offering a practical way to create
ferroelectric-ferromagnets.

Although this constitutes a nice proof of concept, the practical use of
EuTiO$_3$ will be limited by its extremely low N\'eel temperature ($T_N
\simeq$ 5.5~K). In the search of alternative systems realizing the same ideas,
it might be interesting to look for compounds having magnetic ordering
temperatures as high as possible. However, this is limited by the fact that
the difference of energy between FM and AFM states cannot be too different
from the energy gain produced by the ferroelectric distortion. SrMnO$_3$
exhibiting moderate N\'eel temperature (T$_N$ $\simeq$
233-260~K~\cite{SrMnO3-TN1}) and large spin-phonon coupling was identified by
Lee and Rabe~\cite{SrMnO3-Rabe} as a good candidate. They showed that
increasing epitaxial strains (both tensile and compressive) brings the system
into a ferroelectric-ferromagnetic ground-state through a complex sequence of
consecutive phase transitions.

Again such a general strategy might be applied to other classes of
compounds. For instance MnF$_2$~\cite{MnF2-Julien} was shown to exhibit a low
frequency polar mode, slightly softening with temperature and exhibiting a
sizable spin-lattice coupling.


\section{Lattice-mode engineering}
\label{hif}

Beyond strain engineering, alternative strategies can be envisaged to create
new multiferroics.  In this section we will discuss how the coupling of polar
modes with non-polar instabilities can be exploited to produce a polar
ground-state in magnetic compounds.

\subsection{Improper and hybrid improper ferroelectrics}

YMnO$_3$ is a well-known multiferroic, which due to its small tolerance
factor, prefers an hexagonal packing to the cubic perovskite form. At the
structural level, it exhibits a structural phase transition from a
paraelectric $P6_3/mmc$ phase to a ferroelectric $P6_3cm$ ground state,
involving unit cell tripling. As evidenced at the first-principles level by
Fennie and Rabe \cite{YMO-improper}, there is no ferroelectric instability at
the zone-center in the $P6_3/mmc$ phase. The phase transition is produced by
the condensation of an unstable zone-boundary $K_3$ mode that, in turn,
produces the appearance of an additional polar distortion through a coupling
term in the energy of the form :
\begin{equation}
\label{eq:YMnO3}
\mathcal{F} \, \approx \lambda  \,\mathcal{Q}_{K_3}^3\,P
\end{equation}
where $\mathcal{Q}_{K_3}$ is the amplitude of the $K_3$ mode. Such compounds
in which the polarization appears as a slave of another non-polar primary
order parameter with which it couples through a term linear in $P$ is called
an improper ferroelectric \cite{levanyuk1974improper}. In contrast to the
strain coupling in Eq.~\ref{P-Strain} that renormalizes the curvature at the
origin of the energy versus polarization well, the {\it improper} coupling in
Eq.~\ref{eq:YMnO3} induces ferroelectricity by shifting this well as
illustrated by the green curves in Fig.~\ref{f:potential}. Improper
ferroelectrics consequently behave differently than proper ferroelectrics.
Importantly, switching the polarization will necessary require switching the
primary non-polar order parameter (dashed green line in
Fig.~\ref{f:potential}). Also, improper ferroelectrics exhibit distinct
dielectric properties \cite{levanyuk1974improper} and are less sensitive to
depolarizing field issues \cite{sai2009absence,Ph-improper}.

A similar improper behavior was recently predicted theoretically by Varignon
and Ghosez \cite{BMO-improper} in 2H-BaMnO$_3$, in spite of a completely
different atomic arrangement, cation sizes, and Mn valence state.  In simple
cubic perovskites, zone-boundary antiferrodistortive (AFD) instabilities
associated to the rotation of oxygen octahedra are very frequent. However, by
symmetry, these modes cannot couple with the polarization through a term
linear in $P$ to produce improper ferroelectricity~\footnote{Hybrid improper
  ferroelectricity can occur in bulk perovskites but involves additional anti-polar modes~\cite{zhou2013hybrid,PhysRevLett.112.057202}.}. The situation
  is however distinct in layered perovskites~\cite{PhysRevB.70.214111}.

\begin{figure}
\centering
\resizebox{15cm}{!}{\includegraphics{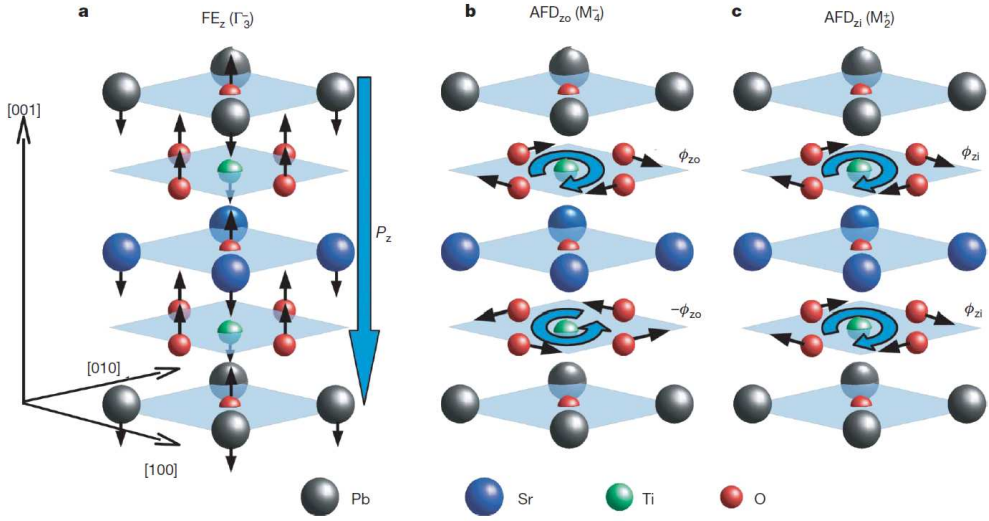}}
\caption{Schematic view of the polar mode (a) and the two AFD motions (b and
  c). Picture taken from Ref.~\cite{PTO-STO}}
\label{f:PTO-STO}
\end{figure}

In 2008, Bousquet {\em et al.} \cite{PTO-STO} reported a new type of improper
ferroelectricity in PbTiO$_3$/SrTiO$_3$ short-period superlattices epitaxially
grown on SrTiO$_3$. From first-principles calculations they revealed that the
ground state of PbTiO$_3$/ SrTiO$_3$ 1/1 superlattices exhibit a complex
ground state combining 2 independent AFD motions and one ferroelectric
distortion sketched in Fig.~\ref{f:PTO-STO} and hereafter referred as $\phi_1$,
$\phi_2$ and $P$ respectively. 
They highlighted that, by symmetry, these 3
modes couple through a trilinear energy term of the form :
\begin{equation}
\label{eq:PTOSTO}
\mathcal{F} \, \approx \lambda  \,\phi_1 \, \phi_2 \,P
\end{equation}
They argued that in systems where $\phi_1$ and $\phi_2$ instabilities
dominate, this term can give rise to ferroelectricity in a way similar to
Eq.~\ref{eq:YMnO3}. In this case, however, two independent non-polar modes of
different symmetry are involved and the term ``hybrid improper
ferroelectricity'' as been coined by Benedek and Fennie~\cite{Ca3Mn2O7-Fennie}
to label such systems. 

Since then, the trilinear coupling of lattice modes in perovskite layered
structures has generated increasing interest. Further investigations have been
since performed on PbTiO$_3$/SrTiO$_3$ superlattices, clarifying the role of
the strain on the relevant AFD motions and the
polarization~\cite{PTOSTO-junquera}. Alternative trilinear couplings have been
obtained~\cite{NaLaMnWO6,Ca3Mn2O7-Fennie,HIF-Rondinelli,MOF2-Picozzi}. Guiding
rules to identify alternative hybrid improper ferroelectrics have been
proposed~\cite{HIF-Rondinelli} and the emergence of ferroelectricity in
rotation-driven ferroelectrics was discussed~\cite{Antiferro-HIF}. Finally,
novel improper couplings have been revealed in alternative A-cation ordered
structures such as the [111]-rocksalt arrangement~\cite{young2014improper}.

\subsection{Toward electric control of the magnetization}

As in regular improper ferroelectrics, the switching of the polarization in
hybrid improper ferroelectrics will necessarily be associated to the switching
of another non-polar mode (either $\phi_1$ or $\phi_2$). It had been suggested
by Bousquet {\em et al}~\cite{PTO-STO} that the intimate link between polar
and AFD motions produced by the trilinear term could be exploited to tune the
magnetoelectric response.

Benedek and Fennie~\cite{Ca3Mn2O7-Fennie} proposed to realize this in
Ca$_3$Mn$_2$O$_7$, a naturally occurring Rudlesden-Popper layered
compound. Using first-principles calculations, they showed that in this system
two AFD motions not only combine together to induce a polar distortion through
Eq. \ref{eq:PTOSTO} but also produce weak ferromagnetism and a linear
magneto-electric effect. They proposed that under appropriate strain
engineering of the energy landscape, it might be possible to realize electric
switching of the magnetization in such systems.

A similar type of hybrid improper ferroelectricity has recently been reported
in alternative magnetic systems such as NaLaMnWO$_6$~\cite{NaLaMnWO6} or
RLaMnNiO$_6$~\cite{RLaNiMnO6} double perovskites, BiFeO$_3$/LaFeO$_3$
superlattices~\cite{BFO-LFO} and even in metal-organic
frameworks~\cite{MOF2-Picozzi}.

Some effort has been devoted recently to rationalize the concept of improper
ferroelectrics from finite electric displacement calculations at
0~K~\cite{Ph-improper}. Still at this stage several important questions remain
open regarding these materials. For example uncertainties remain concerning
their finite temperature properties and in particular the phase transition
mechanism~\cite{PhysRevB.70.214111}: for example are there single or
consecutive phase transitions, or an avalanche
mechanism~\cite{etxebarria2010role}? Another central issue also concerns the
ferroelectric switching path and associated energy barrier. As highlighted
from the first principles calculations in Ref.~\cite{BFO-LFO}, the path with
the lowest energy barrier seems to be quite complex, but compatible with the
reversal of the magnetization. To begin to understand these issues further,
finite temperature molecular dynamics using effective Hamiltonians, along with
further experimental efforts, are likely to play a key role in the future.

Trilinear couplings are not restricted to AFD motions but can alternatively
include Jahn-Teller motions~\cite{MOF2-Picozzi} or anti-polar
motions~\cite{zhou2013hybrid,PhysRevLett.112.057202}. This opens the
possibility to realize similar phenomena in bulk perovskites. In the
identified $Pmc2_1$ phase of highly strained BiFeO$_3$, a trilinear term,
$M_{AP}\,\Phi\,P$, involving an anti-polar mode ($M_{AP}$), one rotation
($\Phi$) and the polarization $P$ was discovered and predicted to allow for an
electrical control of magnetization~\cite{PhysRevLett.112.057202}.

\section{Inducing electronic polarization in magnets through charge, spin and
  orbital ordering}
\label{charge}

In the previous sections, we have primarily discussed how strain and lattice
mode couplings can induce a ferroelectric polarization and how this can help
to design new multiferroics.  A different class of multiferroic exists where
it is the electronic degrees of freedom (charge, spin or orbital) that instead
lowers the symmetry of the system and produce ferroelectricity. The resultant
electronic polarization is expected to be small but can range from
nC.cm$^{-2}$ to several $\mu$C.cm$^{-2}$. The advantages here include
potentially faster polarization switching involving electron rather than ionic
dynamics, and substantially larger magnetoelectric coupling since magnetism
and ferroelectricity can share the same microscopic origin.

First-principles calculations are ideally suited for the study of subtle
microscopic electronic phenomena and elucidating novel electronic
multiferroics has been one of the main focuses within the first-principles
community over the last few years.  Whilst the field of electronic
multiferroics is still in its infancy, exciting new results are constantly
being obtained and first-principles calculations have often played a key
role. Below we very briefly highlight some examples. For further discussions
on this topic we refer the reader to the previous chapter of Picozzi, and to
the recent review of Barone and Yamauchi~\cite{review-Pelec}.

\subsection{Spin ordering induced ferroelectricity}

In some systems, traditionally labelled as type-II multiferroics , the spin
order helps to break the inversion symmetry, which can lead to a polar
ferroelectric state. A prototypical example of this is the orthorhombic
perovskite TbMnO$_3$ where the non-collinear cycloidal-spin structure
generates an electric polarization via the spin-orbit interaction.  Although
the polarization might be thought as purely electronic in nature in such
systems, it was clarified by Malashevich and Vanderbilt \cite{Malashevich-08}
that, in TbMnO$_3$, it has a dominant contribution coming from the subsequent
lattice relaxation.

In the same spirit, it was discovered that ferroelectric polarization can also
be inherent to a collinear E-type antiferromagnetic (AFM-E) order. The AFM-E
ordering consists of up-up-down-down spin ordering, through FM nearest
neighbour and AFM next nearest neighbour interactions, as observed in several
orthorhombic perovskite manganites
RMnO$_3$~\cite{EtypeExt-RMnO3-1,EtypeExt-RMnO3-2} and monoclinic nickelates
RNiO$_3$~\cite{EtypeExt-RNiO3}.  First-principles calculations have been a
very powerful tool in solving the origin of the polarization in these
particular cases. For instance, DFT calculations were able to validate that
AFM-E induces ferroelectricity in one of the prototypical compounds
HoMnO$_3$~\cite{HoMnO3-Picozzi,HoMnO3-Picozzi2} and to explain the relatively
small polarization measured in HoMn$_2$O$_5$ and TbMn$_2$O$_5$, understood as
a cancellation of the contribution coming from the atomic displacements and
the electronic part due to strong electronic
correlations~\cite{HoMn2O5-Annulation,TbMn2O5-Annulation}.  This electronic
polarization has been shown to exist without the need of any lattice
displacements, or spin-orbit interaction, as observed in
YMn$_2$O$_5$~\cite{YMn2O5}.


\subsection{Charge ordering induced ferroelectricity}

Another route to create electronic polarization is achieved by charge ordering
through mixing cations with different valence states. This was first proposed
by Efremov {\em et. al.}~\cite{CO-AMnO3-1} in the doped manganites
(Pr$_{0.4}$Ca$_{0.6}$MnO$_3$ for instance~\cite{CO-AMnO3-2,CO-AMnO3-3}) and
where first principles calculations provided important microscopic
understanding~\cite{Giovannetti}.

A similar mechanism occurs in the well known magnetite Fe$_3$O$_4$ compound.
Magnetite, the first magnetic system ever to be identified, was known for
decades to develop ferroelectricity at low temperature~\cite{Ps-Fe3O4} but its
origin was highly debated. Recent DFT calculations from Picozzi {\em
  et. al.}~\cite{Fe3O4-Picozzi} demonstrated the charge ordering between
Fe$^{2+}$/Fe$^{3+}$ to be responsible for the experimentally measured and
switchable polarization~\cite{Fe3O4-Exp}.

A debate is still ongoing for the similar ferrimagnetic magnetoelectric spinel
LuFe$_2$O$_4$. Indeed, it was proposed that the charge ordering between
Fe$^{2+}$/Fe$^{3+}$ on consecutive triangular Fe bilayers was responsible of
the polarization~\cite{LuFe2O4}, making LuFe$_ 2$O$_4$ the ideal
charge-ordered induced multiferroic. Eventually, a recent joint X-ray plus DFT
study proposed an antiferroelectric charge ordered ground
state~\cite{LuFe2O4-AFE}.  First-principles calculations also predict related
spinels to be good candidates to reach an electronic ferroelectric
multiferroic system, such as in vanadium based spinels
ZnV$_2$O$_4$~\cite{ZnV2O4} or CdV$_2$O$_4$~\cite{CdV2O4} or Fe based systems
which are predicted to show a large magnetoelectric effect~\cite{CaFeO2}.

Following the same spirit, Picozzi {\em et. al.}~\cite{TungstenBronze}
proposed the tetragonal tungsten bronze K$_{0.6}$FeF$_3$ as a prototypical
charge ordered induced ferroelectricity, and a novel playground for
multiferrroicity.  Based on first-principles calculations, various
Fe$^{2+}$/Fe$^{3+}$ charge ordering patterns were found to produce
polarizations with different directions and magnitudes.

\subsection{Orbital ordering induced ferroelectricity}

The third electronic degree of freedom, orbital ordering, is also proposed as
a route to engineer ferroelectricity, however it is commonly linked (induced)
to (by) a charge or spin ordering, and remain more elusive.  For instance,
Ogawa {\em et al} propose on the basis of a joint Second Harmonic Generation
and DFT study that the orbital ordering appearing in some shear-strained half
doped manganite thin films is linked to an off-centering of the
cations~\cite{OO-Fel-ABMnO3}.  Orbital ordering has been predicted to produce
ferroelectricity at the ultra-thin film limit in SrCoO$_3$~\cite{SrCoO3}.
Through partial substitution of Ti$^{4+}$ (3d$^0$) by magnetic vanadium
(3d$^1$) ions in non-magnetic La$_2$Ti$_2$O$_7$, Scarrozza {\em et.  al.}
demonstrated that a multiferroic behaviour is reached with a sizeable
polarization of 4.5 $\mu$C.cm$^{-2}$ at 12\% of doping arising from combined
orbital ordering and V-V dimerisation symmetry breaking.

\section{Interface magnetoelectricity}
\label{metal-isolant}

Whilst recent progress has been made in bulk single-phase magneto-electrics
(including superlattices), a special attention was also devoted to
multi - component / composite systems, where the search is less constrained, and
the magneto-electric effect may be substantially larger. This topic has been
the focus of intensive research over the last decade, with two main strategies
surfacing. The traditional approach uses the strain coupling at an interface
between a piezoelectric and a piezomagnetic \cite{Zheng-04}. The polarization
and magnetization coupling is mediated by strain which can be a longer range
effect penetrating into the bulk of each material. The second strategy is more
fundamental : since any interface/surface of a material breaks spacial
inversion symmetry, if one of the two components is ferromagnetic which
additionally breaks time reversal, a linear magnetoelectric effect can be
expected~\cite{YAMADA}. This section will provide only a brief overview of the
second strategy, which relies on subtle alterations of chemistry, bonding,
structure and electronics at the interface, and hence where first-principles
calculations have proved invaluable. For more comprehensive reviews, we refer
the reader to references~\cite{WANG,VELEV,VAZ,VAZ2,FENNIE}.

Interface systems create two additional theoretical challenges over bulk
systems. Firstly, the task of modelling metal-insulator capacitor systems
under finite electric field. Secondly the problem of band alignment at
metal-insulator interfaces with DFT which famously underestimates band gaps
\cite{Strain-review}. These two challenges have only been considered in recent
years~\cite{STENGEL1,STENGEL2,STENGEL3}. Several DFT codes can now routinely
examine capacitor systems under various electrical boundary conditions, and
many studies are now closely analysing the effect of band-gap correcting
functionals on metal-insulator heterostructures (see for example
references~\cite{DISANTE,CHEN}).

\begin{figure}[h!]
\centering
\resizebox{12cm}{!}{\includegraphics{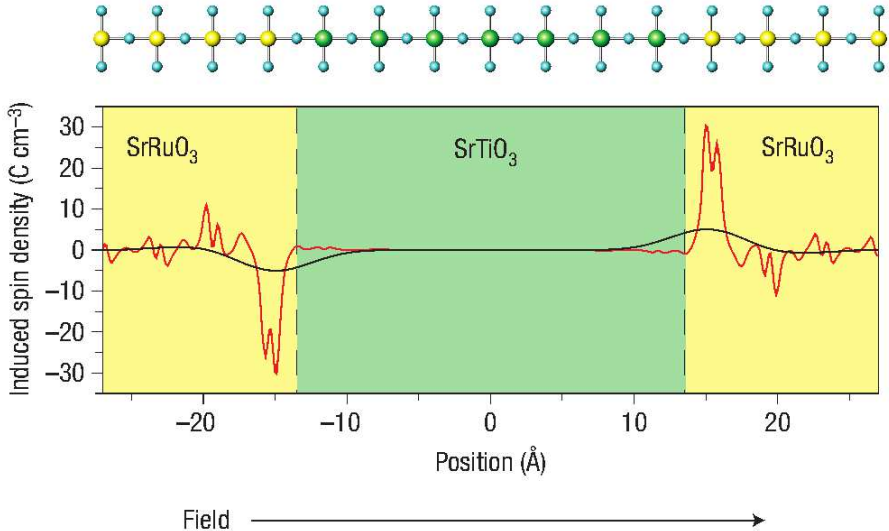}}
\caption{Electric field induced interface magnetization from first principles
  calculations of a SrTiO$_3$-SrRuO$_3$ capacitor~\cite{RONDINELLI}. Red and
  black lines are the calculated planar and macroscopically averaged induced
  magnetizations respectively.  Figure taken from Ref.~\cite{RONDINELLI}.}
\label{f:interface}
\end{figure}

With these recent advances in theoretical methodologies, first principles
calculations are leading the way in the field of magnetoelectric
interfaces~\cite{RAMESH}, not only in the fundamental understanding of
experimental observations, but interestingly in the prediction of new
phenomena. One striking example is the study of Rondinelli {\em et
  al}~\cite{RONDINELLI}, where a novel carrier-mediated magnetoelectric effect
was demonstrated at the SrRuO$_3$/SrTiO$_3$ interface. The effect was argued
to be a universal feature of metallic-ferromagnetic/dielectric interfaces,
where spin-polarized carriers within the metal accumulate or deplete in the
interface region in an attempt to screen the capacitive/bound charges arising
at the interface under an electric field (see figure ~\ref{f:interface}). The
effect was argued to be magnified when the dielectric is replaced with a
ferroelectric, here BaTiO$_3$ (BTO) and if the metal displays a high spin
polarization at the Fermi level, whilst low total magnetization (as in a
half-metallic antiferromagnet)~\cite{RONDINELLI,NIRANJAN2}. First-principles
calculations have observed related carrier-mediated magnetoelectric effects,
though substantially weaker, at magnetic metallic surfaces, such as
SrRuO$_3$~\cite{NIRANJAN2}, Fe$_3$O$_4$~\cite{DUAN2}, and Fe, Co and
Ni~\cite{DUAN3}.

At the Fe/BaTiO$_3$ interface, calculations suggested an induced moment on
interface Ti atoms, whose magnitude depends on the BaTiO$_3$ polarization
direction~\cite{DUAN,FECHNER}. Authors ascribed a mechanism based on local
atomic distortions at the interface. Depending on the polarization direction,
hybridization between Ti $3d$ and Fe $3d$ strengthens or weakens, altering the
moment on Ti. This explanation has recently been challenged within
reference~\cite{STENGEL3} where the authors suggested that the aforementioned
problem related to pathological band-offsets within DFT may be playing a
role. A more recent DFT+U study correcting for the band gap, ascribed the
magnetoelectric coupling as a combination of hybridization and carrier
mediated~\cite{LEE}. Similar hybridisation-driven magnetoelectric effects have
also been predicted from first principles for Fe/PbTiO$_3$~\cite{FECHNER},
Co$_2$MnSi/BaTiO$_3$~\cite{YAMAUCHI}, and
Fe$_3$O$_4$/BaTiO$_3$~\cite{NIRANJAN} interfaces.

First-principles calculations on similar metallic-magnetic/insulator systems
have observed a range of additional fascinating phenomena. Namely the electric
field has been predicted to tune not only the magnitude of the magnetic
moments, but also the magnetic ordering, the magnetic easy axis and the
magnetic N\'{e}el Temperature. Each of these mechanisms are described briefly
below.

The carrier mediated effect described above can be viewed as a local change in
the doping concentration near the interface. In materials such as the doped
manganites (eg La$_{1-x}$Sr$_x$MnO$_3$), the doping concentration plays a
dramatic role on the physical properties of the material, in particular the
magnetic ordering. First principles calculations elucidated such an effect at
an interface between La$_{1-x}$A$_x$MnO$_3$(A=Ca,Sr,Ba) and
BTO~\cite{BURTON}. Depending on the direction of the polarization in
BaTiO$_3$, the ground state magnetic phase was found to be either
ferromagnetic or A-type AFM locally at the
interface~\cite{BURTON,BRISTOWE,CHEN}.

In addition to the electric field induced changes in the magnitude of the
magnetic moment, or the magnetic ordering, first principles calculations have
predicted the possibility of electric-field tuning of the interface/surface
magnetic easy
axis~\cite{DUAN3,DUAN4,NAKAMURA,TSUJIKAWA,ZHANG,NIRANJAN3,LUKASHEV}. The
alteration of the magnetocrystalline anisotropy was argued to arise from field
driven changes in the relative occupations of the $t_{2g}$ orbitals for the
case of Fe/MgO~\cite{NIRANJAN3} and several ferromagnetic metallic
surfaces~\cite{DUAN3,NAKAMURA}. For the Fe/BaTiO$_3$ interface the effect was
instead thought to arise from Ti $3d$-Fe $3d$ hybridisation
modification~\cite{DUAN4}. Recently first principles calculations have
predicted a 180$^{\circ}$ switch in the magnetization at the Fe/PbTiO$_3$
interface through reversal of the polarizaton~\cite{FECHNER2}. This was
produced by utilizing the magnetic interlayer exchange coupling with a capping
FM/nonmagnetic/FM trilayer.

Finally we mention the predicted and demonstrated enhancement of the Curie
temperature in a ferromagnetic-ferroelectric superlattice~\cite{SADOC}. The
effect was argued as a modification of the orbital ordering arising on Mn $3d$
orbitals in La$_{1-x}$Sr$_x$MnO$_3$/BaTiO$_3$ superlattices, which can affect
the strength of the FM double exchange mechanism through enhancing orbital
overlap. The role of the polarization of ferroelectric BaTiO$_3$ here was not
discussed, only the epitaxial strain mismatch. However the calculations
revealed a large asymmetry on Mn interface moments which is an indication of
spin-polarized carrier screening of the ferroelectric polarization.
Ferroelectric distortions have recently been argued to play a role in the
orbital ordering in related PbTiO$_3$-La$_{1-x}$Sr$x$MnO$_3$
superlattices~\cite{CHEN2} through first principles calculations.

\section{Conclusions}

Over the last twenty years, first-principles methods have proven their
effectiveness within the field of multiferroics, both for explaining phenomena
and designing novel materials.  In this chapter, we have summarized the
advances in the theoretical methodologies and the key discoveries for which
the methods have fuelled.  Naturally the multiferroic community has focussed
on the ABO$_3$ perovskites, likely due to historical reasons with the push
from the ferroelectric section.  With the first principles conception of
several design strategies for novel multiferroics, perhaps the future focus
should be the targeting of new promising classes of materials other than
perovskites.  In this respect, the combination of these design strategies,
along with the rapid rise of the so-called
high-throughput~\cite{setyawan2010high,hautier2012computer,Jain2013,bennett2012hexagonal}
first-principles calculations, could pave the way for future multiferroic
design.

\section*{Acknowledgments}
Work supported by ARC project TheMoTherm. Ph. Ghosez acknowledges a research
professorship from the Francqui foundation. E. Bousquet acknowledges FRS-FNRS.

\end{document}